\documentclass[amsmath,amssymb,aps,pra,twocolumn,showpacs]{revtex4-2} 
\usepackage{graphicx}
\usepackage{bm}
\usepackage{hyperref}
\usepackage{color}
\usepackage[normalem]{ulem}

\begin{document}

\title{Beyond the Tavis-Cummings model: revisiting cavity QED with atomic ensembles}

\author{Martin Blaha$^1$}
\email{martin.blaha@hu-berlin.de}
\author{Aisling Johnson$^{2}$}
\altaffiliation{Present address: Vienna Center for Quantum Science and Technology, Faculty of Physics, University of Vienna, 1090 Vienna, Austria.}
\author{Arno Rauschenbeutel$^1$}
\author{J\"urgen Volz$^1$}
\email{juergen.volz@hu-berlin.de}
\affiliation{$^1$ Department of Physics, Humboldt-Universit\"at zu Berlin, 12489 Berlin, Germany,}
\affiliation{$^2$Vienna Center for Quantum Science and Technology, Atominstitut, Technische Universit\"at Wien, 1020 Vienna, Austria.}

\date{\today}

\begin{abstract}
The interaction of an ensemble of $N$ two-level atoms with a single mode electromagnetic field is described by the Tavis-Cummings model. 
There, the collectively enhanced light-matter coupling strength is given by $g_N = \sqrt{N} \bar{g}_1$, where $\bar{g}_1$ is the ensemble-averaged single-atom coupling strength.
Formerly, this model has been employed to describe and to analyze numerous cavity--based experiments. Here, we show that this is only justified if the effective scattering rate into non-cavity modes is negligible compared to the cavity's free-spectral range. In terms of experimental parameters, this requires that the optical depth of the ensemble is low, a condition that is violated in several state-of-the-art experiments. 
We give quantitative conditions for the validity of the Tavis-Cummings model and derive a more general Hamiltonian description that takes into account the cascaded interaction of the photons with all consecutive atoms. We show that the predictions of our model can differ quantitatively and even qualitatively from those obtained with the Tavis-Cummings model.
Finally, we present experimental data, for which the deviation from the predictions of the Tavis-Cummings model is apparent.
Our findings are relevant for all experiments in which optically dense ensembles of quantum emitters are coupled to an optical resonator.
\end{abstract}

\maketitle

\section{Introduction}
\label{introduction}

When strongly coupling a quantum emitter to a high-finesse optical resonator, one can control light-matter interaction at the fundamental level of individual photons. Such coupled emitter--resonator systems have proven successful for studying and harvesting cavity quantum electrodynamics (cQED) effects, ranging from the implementation of textbook models \cite{thompson_observation_1992, Brune1996} to quantum-enhanced protocols with potential applications in quantum technology \cite{Reiserer2015}. Recently, an increasing number of experiments go beyond cQED with single emitters and make use of the collective interaction of an ensemble of, e.g., atoms with the resonator mode to study complex many-body problems. These experiments encompass cold atomic clouds \cite{tuchman_normal-mode_2006, Johnson2019} or Bose-Einstein condensates \cite{brennecke_cavity_2007, colombe_strong_2007, vaidya_tunable-range_2018} that are coupled to one or several resonator modes with the aim of exploring phase transitions or novel regimes of cQED such as superstrong coupling \cite{meiser_superstrong_2006, Johnson2019}. The interaction of atomic ensembles with optical cavities has furthermore been proposed for metrology applications \cite{tuchman_normal-mode_2006} or, in an integrated on-chip configuration, as a platform for quantum information processing \cite{ripka_room-temperature_2018, hu_cavity-enhanced_2019}. A common aspect of many of these experiments is the realization of a large light-matter coupling strength based on the coherently enhanced collective interaction of many atoms with the cavity field. According to the Tavis-Cummings model \cite{jaynes_comparison_1963, tavis_exact_1968}, which we briefly recall in section \ref{sec.TC} below, the dynamics of the coupled system in the low excitation regime is identical to single-atom cQED but with a coupling strength $g_N = \sqrt{N} \bar{g}_1$ \cite{jaynes_comparison_1963, tavis_exact_1968, tavis1968exact}, where $N$ is the number of atoms that collectively interact with the resonator mode and $\bar{g}_1$ is the ensemble-averaged single atom-resonator coupling strength.
Here, we study the limits of this approach and show that the standard Tavis-Cummings model breaks down for larger atom numbers (section \ref{sec.breakdown}) under conditions reached in many experiments. In order to go beyond the Tavis-Cummings model, we develop a real-space description of the coupled system (section \ref{sec.wgtocqed}) that considers the successive interaction of the $N$ atoms with the propagating cavity field. We formulate a general Hamiltonian (section \ref{sec.waveguide}), which is valid in a larger parameter range than the Tavis-Cummings model. We then derive an analytical solution for the stationary state of an ensemble of atoms interacting with the modes of a ring resonator (section \ref{sec.solutions}) and compare the predictions of our model to the Tavis-Cummings model as well as to a generalization of the Tavis-Cummings model that accounts for more than one cavity mode (section {\ref{sec.comparison}}). Finally in section \ref{sec.experiment}, we apply our model to an experiment for which deviations from the Tavis-Cummings predictions are expected. We show that our model correctly describes the experiment where an ensemble of atoms is ``superstrongly'' coupled \cite{meiser_superstrong_2006} to a 30-m long ring resonator.

\section{The Tavis-Cummings model}
\label{sec.TC}

In the framework of the rotating wave approximation, the interaction of a single two-level quantum emitter, e.g., a single atom, with a single electromagnetic field mode is described by the Jaynes-Cummings (JC) model \cite{jaynes_comparison_1963, tavis_exact_1968, tavis1968exact}. The Hamiltonian reads: 
\begin{equation}
	\frac{\hat{H}_{JC}}{\hbar} = \omega_a \hat{\sigma}^+ \hat{\sigma}^- + \omega_c \hat{a}^{\dagger} \hat{a} + g_1\left(\hat{a}^{\dagger} \hat{\sigma}^- + \hat{a} \hat{\sigma}^+\right).
\end{equation}
Here, $\hat{\sigma}^+$ ($\hat{\sigma}^-$) denotes the atomic raising (lowering) operator and $\hat{a}^{\dagger}$ ($\hat{a}$) is the photon creation (annihilation) operator. $\omega_a$ and $\omega_c$ are the atomic and cavity resonance frequencies, respectively, and $g_1$ is the single atom--single photon coupling strength which, without loss of generality, we assume to be real and positive.

Within the single excitation manifold of the \{atom + cavity\} system, the coupling of the atom to the resonator results in two new eigenstates with eigenenergies
\begin{equation}
E_{JC} = \hbar\omega_c -\frac{\hbar}{2}\left(\Delta_{ca} \pm \sqrt{4g_1^2+\Delta_{ca}^2}\right)
\end{equation}
where $\Delta_{ca}=\omega_c-\omega_a$ is the cavity--atom detuning. For $\Delta_{ca}=0$, this leads to the vacuum Rabi splitting of the unperturbed resonance into two split resonances separated in frequency by $ 2 g_1$.

When describing an experimental \{atom + cavity\} system, we have to account for losses. 
Two loss channels are typically considered: the spontaneous emission rate of the atom into all free-space modes, $ \gamma_l $, and the cavities field decay rate, $ \kappa_{0} $, quantifying intra-cavity losses. These loss rates can be taken into account by introducing complex emitter and cavity resonance frequencies $ \tilde{\omega}_a = (\omega_{a}-i\gamma_l) $ and $ \tilde{\omega}_c = (\omega_c-i\kappa_{0}) $, respectively.
For coupling the \{atom + cavity\} system with an external probe field, we the Hamiltonian has to be extended by the operator $ \hat{U}_\textrm{probe}$ which, e.g., for the case of a coherent probe field can be described as $\hat{U}_\textrm{probe} = i\sqrt{2\kappa_{\text{ext}}}\eta(\hat{a}^\dagger+\hat{a})$, where $ \kappa_{\text{ext}} $ is the coupling strength between the fields in and outside of the cavity and $ \eta $ is the amplitude of the probe field. Note that for the calculation in this manuscript we use the more general operator $\hat{U}_\textrm{probe} $ in \ref{app.strong_coupling}.
The resulting Hamiltonian, which now includes loss rates and probing, is given by
\begin{equation}\label{eq.JC_loss}
\frac{\hat{H}_{JC'}}{\hbar} = \tilde{\omega}_a \hat{\sigma}^+ \hat{\sigma}^- + \tilde{\omega}_c \hat{a}^{\dagger} \hat{a} + g_1\left(\hat{a}^{\dagger} \hat{\sigma}^- + \hat{a} \hat{\sigma}^+\right) + \hat{U}_\textrm{probe}.
\end{equation}
A key figure that quantifies the performance of such an atom-resonator system is the so-called cooperativity of the coupled system
\begin{equation}
C=\frac{g_1^2}{2\kappa_0\gamma_l},
\end{equation}
where most experiments and applications aim for the regime $C \gg 1$. In this regime, the coupled system operates closely to the ideal, lossless system.

If more than one atom is coupled to the resonator mode, the Jaynes-Cummings description has to be extended, yielding the Tavis-Cummings (TC) model \cite{tavis_exact_1968}. Restricting the Hilbert space of the atoms to the sub-space spanned by the fully symmetric Dicke states, and considering at most one excitation in the system, the collective atomic excitation and annihilation operators are
\begin{equation}
	\hat{S}^+=\frac{1}{\sqrt{N}}\sum_{n=1}^N \hat{\sigma}^+_n \;\; \textrm{and } \; \hat{S}^-=\frac{1}{\sqrt{N}}\sum_{n=1}^N \hat{\sigma}^-_n.
\end{equation}
Using these operators, the TC Hamiltonian in the low excitation limit reads
\begin{equation}
	\frac{\hat{H}_{TC}}{\hbar} = \omega_a \hat{S}^+ \hat{S}^- + \omega_c \hat{a}^{\dagger} \hat{a} + g_N \left(\hat{a}^{\dagger} \hat{S}^- +  \hat{a} \hat{S}^+\right). \label{eq.TC}
\end{equation}
A Hamiltonian describing the probed and lossy system can then be defined analogously to Eq.~(\ref{eq.JC_loss}).
From Eq.~(\ref{eq.TC}), it becomes apparent that the interaction between the atomic ensemble and the resonator mode is formally the same as for a single atom, but with a collectively increased coupling strength, $g_{N}=\sqrt{N} g_1$. The $N$ emitters thus collectively behave as a superatom with a $\sqrt{N}$-fold increased coupling strength, thereby providing a straightforward strategy for enhancing light-matter coupling in experiments. Note that each atom coupled to the resonator mode can in principle have a different coupling strength, $ g_{1,n} $, e.g., due to the spatial variation of the cavity field. In this case, the Tavis-Cummings-Hamiltonian Eq.~(\ref{eq.TC}) still applies when defining the collective coupling strength as
\begin{equation}
g_N=\big(\sum_{n=1}^N g_{1,n}^2\big)^{1/2}\equiv\sqrt{N}\bar{g}_1,
\end{equation}
where $\bar{g}_1=(\sum g_{1,n}^2/N)^{1/2}$ is the root-mean-square of the individual atomic coupling strengths.

\section{Limits of the Jaynes-Cummings and the Tavis-Cummings models}
\label{sec.breakdown}

We now discuss the conditions under which the Jaynes-Cummings and the Tavis-Cummings approaches are valid. 
An implicit assumption in the above models is that all relevant rates and frequency scales are small compared to the free spectral range $\nu_{\text{FSR}}$ (i.e., the inverse of the photon round trip time) of the resonator. However, two quantities can potentially violate this condition when the atom number $N$ increases: the collective coupling strength, $g_N$, and the atom-induced photon loss rate of the system, $g_N^2/\gamma_l$, see Appendix~\ref{app.strong_coupling}. The JC or TC Hamiltonian is therefore only a valid description of the system if the  inequalities 
\begin{eqnarray}
	\nu_{\text{FSR}}&\gg&  g_N \label{eq.cond1a}\\
	\nu_{\text{FSR}}&\gg&  \frac{g_N^2}{\gamma_l}\label{eq.cond1b}
\end{eqnarray}
are fulfilled. Violating condition (\ref{eq.cond1a}) implies that the collective coupling strength reaches or exceeds the free spectral range of the resonator, thereby entering the so-called superstrong coupling regime of cQED \cite{meiser_superstrong_2006, Johnson2019}. This condition will typically only break down in experiments that were designed to explore this regime. Furthermore, the superstrong coupling regime can straightforwardly be included in the above description by considering the coupling of the atoms to many cavity modes. In doing so, one obtains the multimode Tavis-Cummings Hamiltonian, which reads 
\begin{equation}\label{eq.TC_multimode}
\frac{\hat{H}_{TCmm}}{\hbar} = \omega_{a}\hat{S}^+\hat{S}^-+ \sum_j\left[\omega_j\hat{a}^\dagger_j\hat{a}_j + g_{N,j}\left(\hat{a}^\dagger_j\hat{S}^-+\hat{a}_j \hat{S}^+\right)\right],
\end{equation}
where $\hat{a}^\dagger_j $ ($\hat{a}_j $) creates (annihilates) a photon in the $j$th mode of the resonator with frequency $ \omega_j$ and $g_{N,j}$ is the coupling strength between the emitter and resonator mode $j$.

Condition (\ref{eq.cond1b}) is more frequently violated in experiments, as is apparent from table \ref{table}, where we compare the parameters of different cQED experiments.
Contrary to condition (\ref{eq.cond1a}), the JC or TC model cannot be extended to account for the violation of (\ref{eq.cond1b}), because when $ g_N^2/\gamma_l $ exceeds the free spectral range, the atom-field interaction can no longer be treated as instantaneous. Consequently, one requires a description which considers the successive interaction of the field with each atom. In the following, we will establish a model that is based on this approach.

\begin{center}
	\begin{table*}[t!]
		\begin{tabular}{| r | c | c | c | c | c | }
			\hline
			Reference & $g_N$ & $\gamma_l\approx\gamma$ & $\nu_{\text{FSR}}$ & $g_N/{\nu_{\text{FSR}}}$ & $g_N^2/({\gamma_l\nu_{\text{FSR}}})$ \\
			\hline
			\hline
			Lee et al. \cite{lee_many-atomcavity_2014} & $2\pi\times 44.9$ MHz & $2\pi\times$3 MHz & 1.4 GHz & 0.2 & \textbf{3} \\
			\hline
			Johnson et al. \cite{Johnson2019} & $2\pi\times 9.2$ MHz & $2\pi\times$2.61 MHz & 7.1 MHz & \textbf{8.1} & \textbf{29} \\
			\hline
			Brennecke et al. \cite{brennecke_cavity_2007} & $2\pi\times$3.5 GHz & $2\pi\times$3 MHz & 850 GHz & 0.026 & \textbf{30} \\
			\hline
			Vaidya et al. \cite{vaidya_tunable-range_2018} & $2\pi\times$464.9 MHz & $2\pi\times$3 MHz & 15 GHz & 0.195 & \textbf{30} \\
			\hline
			Jiang et al. \cite{jiang_intracavity_2019} & $2\pi\times$313 MHz & $2\pi\times$2.87 MHz & 5.3 GHz & 0.374 & \textbf{41} \\
			\hline
			Colombe et al. \cite{colombe_strong_2007} & $2\pi\times$12 GHz & $2\pi\times$3 MHz & 3.9 THz & 0.019&\textbf{77}\\
			\hline
		\end{tabular}
		\caption{A list of experiments that violate conditions \ref{eq.cond1a} or \ref{eq.cond1b}. Note that for the last column, we assumed no collectively enhanced emission of the ensemble into free space.}\label{table}
	\end{table*}
\end{center}

\section{Cavity and waveguide QED}
\label{sec.wgtocqed}
\begin{figure}[h!]
	\includegraphics[width=0.9\columnwidth]{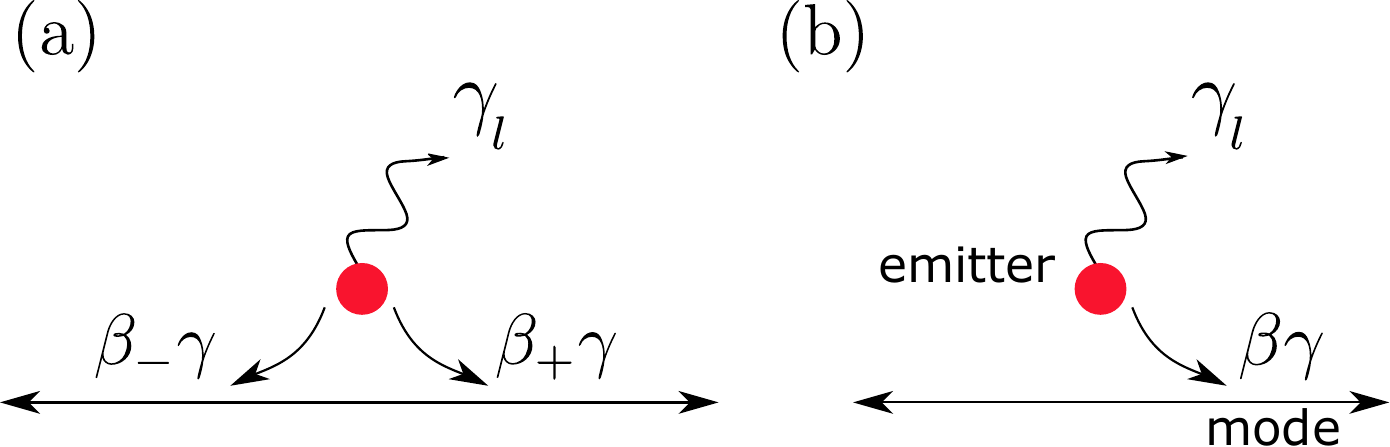}	
	\caption{A single emitter interacting with an optical mode. (a) In principle, an excited emitter can decay through three decay channels: forwards or backwards into the mode with rates $\beta_+\gamma$ or $\beta_-\gamma$ respectively, or into free-space with the rate $\gamma_l=(1-\beta) \gamma$, where $\beta\equiv\beta_++\beta_-$, and, in general, $\beta_+\neq\beta_-$. (b) For the scenario of chiral coupling assumed in chapter \ref{sec.solutions}, $ \beta_- = 0 $ and $ \beta_+ = \beta $ \cite{lodahl_chiral_2017}.}
	\label{fig.theosketch_singleat}
\end{figure}

At first, we will introduce  typical waveguide QED parameters and relate them to the cQED quantities used in section \ref{sec.TC}. For this, we consider a quantum emitter coupled to a propagating optical mode, as sketched in Fig.~\ref{fig.theosketch_singleat}.
The emitter-mode coupling strength for this configuration is typically characterized by the $\beta$-factor, defined as the ratio of the spontaneous emission rate of the emitter into the propagating mode, $\gamma_{mode}$, and the emission rate into all modes, $\gamma=\gamma_l+\gamma_{mode}$. In general, as has recently been demonstrated \cite{lodahl_chiral_2017}, this emitter-mode coupling is not necessarily symmetric with respect to the forward and backwards propagating modes. Therefore, two coupling constants must be introduced, denoted as $\beta_+$ for the forward and $\beta_-$ for the backward propagating modes, with $\beta\equiv\beta_+ + \beta_-$. 
If the waveguide is now transformed into a Fabry-P\'erot or ring resonator with a free spectral range $\nu_{\textrm{FSR}}$ (either by terminating the waveguide with mirrors or by closing the waveguide to form a ring), the vacuum Rabi-frequency of the coupled emitter-resonator system is given by (see Appendix~\ref{app.g_to_beta})
\begin{equation}\label{eq.gvsfsr0}
	g_1^2=4\beta\gamma\nu_{\text{FSR}}
\end{equation}
for a Fabry-P\'erot cavity with the emitter placed at the anti-nodes of the standing wave in the cavity, and by
\begin{equation} \label{eq.gvsfsr}
g_1^2=2\beta\gamma\nu_{\text{FSR}}
\end{equation} 
for an emitter that is chirally coupled to a ring resonator.
These equations give a direct relationship between characteristic cQED and waveguide QED parameters and illustrate that the emitter-resonator coupling strength is given by the geometric mean of the decay rate into the resonator mode, $\beta\gamma$, and the free spectral range, $ \nu_{\text{FSR}} $.
From this, we also obtain the largest possible single emitter-resonator coupling strength for a given resonator length $g_{\text{max}}=2\sqrt{\gamma\nu_{\text{FSR}}}$. 
Interestingly, for a given free spectral range, this maximum coupling strength is limited even though the cooperativity of the system approaches infinity for a perfectly coupled emitter, i.e., $\gamma_l\rightarrow0$.
We note that for free-space cavities, one can only approach this upper limit by employing cavity mirrors that cover almost the full solid angle such that emission into radiative modes can be neglected, i.e., $\gamma_l\approx0$.

\section{Collective coupling in the Dicke and timed-Dicke model}
Concerning the nature of the collective interaction of many emitters, e.g., an ensemble of identical atoms, with the cavity mode, we have to distinguish two basic cases: either the distance between any two emitters is much smaller than $\lambda/2\pi$, or the inter-emitter distance is much larger than $\lambda/2\pi$. The two cases are typically described in the framework of Dicke states \cite{Dicke} and timed-Dicke states \cite{scully}, respectively. Here, $\lambda$ refers to the wavelength of the light that probes the coupled system. 

For the situation where all emitters are separated by less than $\lambda/2\pi$, the cavity field will excite the atoms into the superradiant Dicke state and, assuming the low excitation limit, one excitation is shared between all atoms. Due to the small separations, the emissions of the individual emitters add up constructively for all emission directions. Thus, the ensemble can be described as an effective ``superatom'' with a collectively enhanced dipole moment or, equivalently, a collectively enhanced decay rate $\gamma_N$. When the near field interaction between the atoms is neglected, the latter is given by $\gamma_N=N \gamma$. According to Eq.~(\ref{eq.gvsfsr0}) and Eq.~(\ref{eq.gvsfsr}), this results in a collectively enhanced coupling strength to the resonator of $g_N=\sqrt{\gamma_N/\gamma}\,\,\bar{g}_1$. For this situation, the Tavis-Cummings model typically applies: As $g_N^2$ and $\gamma_{l,N}=(1-\beta)\gamma_N$ are both proportional to $\gamma_N$, their ratio is independent of $N$. Thus, if Eq.~(\ref{eq.cond1b}) is fulfilled for a single emitter, it is also fulfilled by the ensemble. However, experimentally, is hard to confine the emitters to such a small volume. 
And even if this is achieved, the collective interaction of the emitters with the cavity will not increase the cooperativity:
Given that $g_N^2\propto N$ and $\gamma_N\propto N$, see above, the collective cooperativity of the coupled system, $C_N= g_N^2/(2\, \gamma_{l,N}\,\kappa_0)$, is independent of $N$.

The situation is different for the case of large emitter-emitter distances. In the low-excitation regime, the collective emitter-light interaction again gives rise to a state with one shared atomic excitation, which is superradiant with respect to the cavity mode and exhibits a collectively enhanced coupling strength, $g_N=\sqrt{N}\bar{g}_1$. However, in contrast to the situation above, the amplitudes for emission into free space now add up incoherently and the emission rate remains unchanged, $\gamma_{l,N}=\gamma_l$. Consequently, for large emitter-emitter distances, the ensemble-resonator cooperativity will increase linearly with $N$. This is the situation that is aimed for and realized in typical experiments. However, in this case, Eq.~(\ref{eq.cond1b}) will depend on the atom number, and the equation will be violated from a certain emitter number onward.
Using $ g_N = \sqrt{N}\bar{g}_1 $ and Eq.~(\ref{eq.gvsfsr}), we can reformulate Eqs.~(\ref{eq.cond1a}) and (\ref{eq.cond1b}) and obtain for a ring resonator with chiral coupling
\begin{eqnarray}
	\beta N &\ll& \frac{1}{2}\frac{\nu_{\text{FSR}}}{\gamma}\label{eq.cond_beta_mmsc} \\
	\beta N &\ll& \frac{1}{2}(1-\beta),\label{eq.cond_beta}
\end{eqnarray}
where $\beta=\sum \beta_n/N$ is the arithmetic mean of $\beta_n$ of the individual emitters.
For most experiments, $\nu_{\text{FSR}}\gg\gamma$ and $\beta\ll1$. Under these assumptions, condition (\ref{eq.cond_beta}) is always violated first when $N$ increases and takes the simple form
\begin{equation}
	\beta N\ll 1/2. \label{eq.cond_beta2}
\end{equation}
For the case of a Fabry-P\'erot resonator we obtain the same expressions as in Eqs.~(\ref{eq.cond_beta_mmsc})-(\ref{eq.cond_beta2}) but with an additional factor $1/2$ on the rhs.
When considering the interaction of a spatially extended ensemble in a typical cavity QED setup, the assumptions which underlie the TC model therefore no longer hold if $\beta N$ becomes large. In this case, the modification of the resonator field in a single roundtrip becomes significant because the single-pass optical depth of the atomic ensemble, given by $OD\approx4\beta_+ N$ (for $\beta\ll1$), is no longer small compared to 1.
In this case the local strength of the cavity field depends on its interaction with the preceding atoms, and atoms further along the direction of propagation experience a weaker field.

\section{Hamiltonian including cascaded atom-light interaction}
\label{sec.waveguide}

\begin{figure}
	\centering
	\includegraphics[width=0.5\columnwidth]{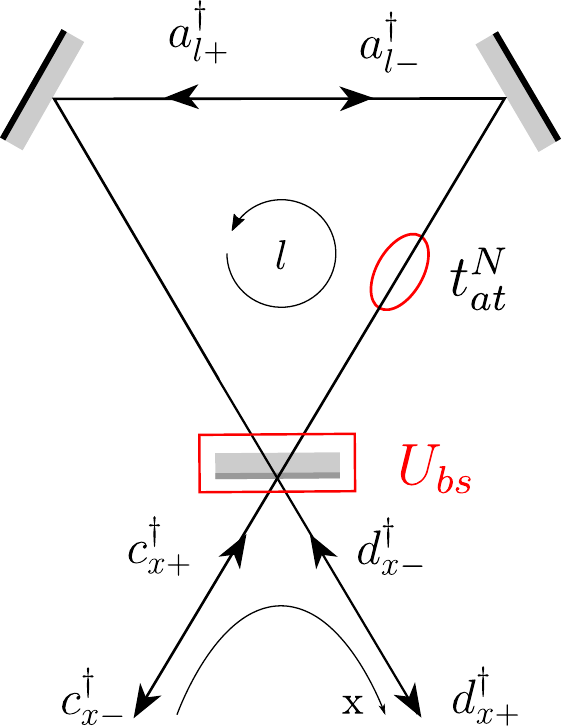}
	\caption{Sketch of the system under consideration featuring the notations introduced in the main text. }
	\label{fig.theosketch}
\end{figure}

In order to derive a Hamiltonian that accounts for the position-dependent resonator field, we follow the formalism in \cite{Shen2009}  and treat the resonator mode as a propagating wave that consecutively interacts with two-level atoms. More specifically, we consider the mode of a ring resonator of length $L$ that is coupled to an ensemble of $N$ atoms and include probing and loss, see Fig.~\ref{fig.theosketch}. The Hamiltonian of this system is given by
\begin{equation}
\label{eq.hamiltonian}
\begin{aligned}
&\hat H/\hbar=\hat{U}_{c}+\\
&\int_{-\infty}^{0}dx \left[ \hat{c}_{x_+}^{\dagger}\left(\omega-i v_{g} \partial_x \right) \hat{c}_{x_+} +\hat{c}_{x_-}^{\dagger}\left(\omega+i v_{g} \partial_x\right) \hat{c}_{x_-}\right] +\\
&\int_{0}^{+\infty} dx \left[\hat{d}_{x_+}^{\dagger}\left(\omega-i v_{g} \partial_x\right) \hat{d}_{x_+} +\hat{d}_{x_-}^{\dagger}\left(\omega+i v_{g} \partial_x\right) \hat{d}_{x_-}\right] + \\
&\int_{0}^{L} dl\Big\lbrace\left[\hat{a}^{\dagger}_{l_+}\left(\omega+i v_{c} {\partial_l}\right) \hat{a}_{l_+}+\hat{a}_{l_-}^{\dagger}\left(\omega-i v_{c} \partial_l\right) \hat{a}_{l_-}\right]+\\
&\qquad\sum_{n=1}^{N}\delta(l-l_n)\Big[\hat{\sigma}_{n}^{+} \hat{\sigma}_{n}^{-}\left(\omega_{\mathrm{a}}-i\gamma_l\right)+\\
&\qquad V_{n,+}\left(\hat{a}_{l_+}^{\dagger}\hat{\sigma}_{n}^{-}+\hat{a}_{l_+}\hat{\sigma}_{n}^{+}\right)+
V_{n,-}\left(\hat{a}_{l_-}^{\dagger}\hat{\sigma}_{n}^{-}+\hat{a}_{l_-}\hat{\sigma}_{n}^{+}\right)\Big]\Big\rbrace.
\end{aligned}
\end{equation}
Here, the intra-resonator field at position $l$ is described by the photon creation and annihilation operators, $\hat{a}_{l_+}^{\dagger}$ and $\hat{a}_{l_+}$ ($\hat{a}_{l_-}^{\dagger}$ and $\hat{a}_{l_-}$), for the counterclockwise (clockwise) propagating mode, respectively, with $l\in[0,L[$. $\hat{\sigma}^+_{n}$ ($\hat{\sigma}^-_{n}$) is the raising (lowering) operator for atom $n$ at position $l_n$, $v_g$ ($v_c$) is the group velocity of light outside (inside) the resonator,  $\omega_{\text{a}}$ is the atomic resonance frequency, and $\partial_x=\partial/\partial x$. The coupling strength of the $n$th atom to the clockwise and counterclockwise propagating resonator mode is given by $V_{n_\pm}=\sqrt{2v_c\beta_{n\pm}\gamma}$.
Here, we assumed that the dispersion relation of the probe field with frequency $\omega_k$ and wavenumber $k$ is approximately linear in the frequency range of interest around the center frequency, i.e., $\omega_{k,\pm} = \omega \pm v_{g,c} k$. The resonator is probed via the incoming and outgoing free-space fields that are denoted by the creation (annihilation) operators $\hat{c}_{x_+}^\dagger,\hat{d}_{x_-}^\dagger$ and 
$\hat{d}_{x_+}^\dagger,\hat{c}_{x_-}^\dagger$ ($\hat{c}_{x_+,}\hat{d}_{x_-}$ and $\hat{d}_{x_+},\hat{c}_{x_-}$), respectively. These modes are coupled to the resonator modes at the incoupling mirror, whose effect is given by the beamsplitter matrix
\begin{eqnarray}
\hat{U}_c&=& \sum\limits_{k=\pm}\Big[t_{\text{rt}}\left(i v_c r \hat{a}_{L_k}\hat{a}^\dagger_{0_k}+\sqrt{v_gv_c}t\hat{a}_{L_k}\hat{d}^\dagger_{0_k}\right) + \nonumber\\
&&\qquad i v_gr\hat{c}_{0_k}\hat{d}_{0_k}^\dagger+ \sqrt{v_cv_g}t(\hat{c}_{0_k}\hat{a}^\dagger_{0_k})\Big],
\end{eqnarray}
where $r$ and $t$ are the amplitude reflection and transmission coefficients of the mirror, which fulfill $r^2+t^2=1$.
Note that, just like for the JC and TC Hamiltonians, the Hamiltonian in Eq.~(\ref{eq.hamiltonian}) above is non-Hermitian because of the term $-i\gamma_l$, which accounts for photon loss by atom-induced scattering out of the resonator and because of the factor $t_{\text{rt}}\leq1$ in the definition of $U_c$, which accounts for resonator roundtrip losses. Furthermore, as discussed above, our approach assumes no collectively enhanced emission into free space. With these definitions, the coefficients $t_\text{rt}$ and $r$ are related to the intrinsic resonator loss rate, $\kappa_0$, and the in-coupling rate, $\kappa_{\text{ext}}$, via $t_{\text{rt}} = \sqrt{1-2\kappa_0/\nu_{\text{FSR}}}$ and $r= \sqrt{1-2\kappa_{\text{ext}}/\nu_{\text{FSR}}}$. 

This description models the successive interaction of the propagating field with $N$ atoms and, thus, naturally considers the modification of the field upon interaction with each individual atom. Moreover, the model also comprises the possible interaction with several longitudinal cavity modes, which occurs in the superstrong coupling regime. In the next section, we will derive an analytical solution for the steady state in the low excitation limit and discuss how the predictions of this model differ from that of the TC model and its generalization to more than one mode.

\section{Solutions in the low excitation limit}
\label{sec.solutions}
In the following, we want to discuss the differences between our model and the TC description. Thus, from now on, we assume perfect chiral emitter-resonator coupling \cite{lodahl_chiral_2017} which allows us to obtain a simple analytical steady-state solution for the Hamiltonian in Eq.~(\ref{eq.hamiltonian}). This means we assume that all atoms interact solely with the forward propagating ($+$ direction) mode via $V_{n,+}$ and set $V_{n-}=0$ in Eq.~(\ref{eq.hamiltonian}). Consequently, we get $\beta_{+,n} = \beta_n$ and $\beta_{-,n}=0$.
Details of the derivation can be found in Appendix~\ref{app.solution}. When probing the resonator along the positive $x$-direction, the power reflection at the input mirror ($c^\dagger\rightarrow d^\dagger$) reads
\begin{equation}\label{eq.solution}
R = \left|\frac{e^{-i\Delta_c/\nu_{\text{FSR}}}t_{\text{rt}}t_{N}-r}{e^{-i\Delta_c/\nu_{\text{FSR}}}t_{\text{rt}}t_{N} r-1}\right|^2, 
\end{equation}
where
\begin{equation} 
t_N = \prod_{n=1}^N t_n
\label{eq.t_N}
\end{equation}
and $t_n$ is the transmission past atom $n$, which is given by (see Appendix~\ref{app.solution})
\begin{equation}
t_{n} = 1-\frac{2\gamma\beta_{n}}{\gamma+i\Delta_a}.
\end{equation}
Here, $\Delta_a = \omega_a - \omega$ and $\Delta_c=\omega_c - \omega$ are the probe-atom and probe-resonator detuning, respectively, and $\omega_c$ is the resonator resonance frequency closest to $\omega_a$.
For small $\beta_n$ or small variations between the individual $\beta_{n}$, we can approximate $t_N$ in Eq.~(\ref{eq.t_N}) by (see Appendix~\ref{app.solution})
\begin{equation}
t_{N} = \left(1-\frac{2\gamma\beta}{\gamma+i\Delta_a}\right)^N, 
\label{eq.tAtom}
\end{equation}
with the mean atom-mode coupling $\beta=\sum \beta_{n}/N$. 
Introducing the complex-valued function
\begin{eqnarray}
\phi(\Delta_a,\Delta_c)&=&-\frac{\Delta_c}{\nu_{\text{FSR}}}+\arg(t_N) -i \ln|t_N t_\textrm{rt}r|\\
&\underbrace{\approx}_{\beta\ll1} & -\frac{\Delta_c}{\nu_{\text{FSR}}}+ \frac{2 \beta N \gamma}{\Delta_a-i\gamma}-i \ln|t_\textrm{rt}r|,
\label{eq.definitionPhi}
\end{eqnarray}
we can simplify Eq.~(\ref{eq.solution}) and obtain
\begin{equation}
	R = \left|\frac{e^{i\phi(\Delta_a,\Delta_c)}/r-r}{e^{i\phi(\Delta_a,\Delta_c)}- 1}\right|^2,
	\label{eq.highODsolution}
\end{equation}
With this definition, $ \mathrm{Re}\left\{ \phi(\Delta_a,\Delta_c) \right\} $ corresponds to the phase shift the light acquires in a single roundtrip in the resonator and $ \mathrm{Im}\left\{ \phi(\Delta_a,\Delta_c) \right\} $ denotes the total roundtrip loss.

We note that, even if the system does not exhibit chiral atom-light coupling, Eqs.~(\ref{eq.solution})--(\ref{eq.tAtom}) still apply provided that the resonator is probed from one direction, that $ \beta \ll 1 $, and that a large number of atoms is coupled to the resonator mode at random positions. Under these conditions, collective enhancement only occurs for the direction in which the resonator is probed, and the residual light scattered by the atoms into the counter-propagating resonator mode \cite{reitz2014backscattering} can be treated as a small loss and does not affect the forward transmission significantly.

Equation~(\ref{eq.solution}) or (\ref{eq.highODsolution}) holds for any combination of cavity parameters. For illustration, we consider two limiting cases. First, when conditions (\ref{eq.cond1a}) and (\ref{eq.cond1b}) are both fulfilled, the atoms effectively only interact with the single resonator mode closest to atomic resonance and the effect of other resonator modes can be neglected. For small detunings ($\Delta_c\ll\nu_{\text{FSR}}$) Eq.~(\ref{eq.solution}) can then be simplified, see Appendix~\ref{app.comptransm}, yielding
\begin{equation}
R = \left|\frac{g_N^2+(\gamma_l+i\Delta_a)\left(\kappa_{0}-\kappa_{\text{ext}}+i\Delta_c\right)}{g_N^2+(\gamma_l+i \Delta_a)\left(\kappa_{0}+\kappa_{\text{ext}}+i\Delta_c\right)}\right|^2,
\label{eq.TCtransmission}
\end{equation}
where $g_N=\sqrt{2N\beta\gamma\nu_{\text{FSR}}}$ is the ensemble-resonator coupling strength. Equation~(\ref{eq.TCtransmission}) is identical to the predictions of the driven JC and TC model \cite{Carmichael2009}, see Appendix~\ref{app.strong_coupling}.
Second, for $r\rightarrow 0$, i.e., in the waveguide limit where light only takes a single roundtrip in the resonator, Eq.~(\ref{eq.solution}) simplifies to $ t_{\text{rt}}^2 |t_{N}|^{2}$. This expression corresponds to a saturated Lorentzian and is the well-known expression for the transmission spectrum of an optically dense atomic ensemble \cite{Sague2007, Vetsch2010}.

We note that our approach can also be used to derive the solution for the case where the light-atom interaction is described in the Dicke picture. In this case, the power reflection is still described by Eqs.~(\ref{eq.solution}) and (\ref{eq.highODsolution}), but the ensemble transmission $ t_N $ has to be replaced by its counterpart for the Dicke case
\begin{equation}
t_{N,D}=1-\frac{2\gamma_N\beta}{\gamma_N+i\Delta_a}
\label{eq.tNDicke}
\end{equation}
to account for the collectively enhanced decay rate $ \gamma_N $.

\section{Beyond Tavis-Cummings Physics}
\label{sec.comparison}
In the following, we study the predictions of the TC model and its generalization to more than one mode when its validity conditions (\ref{eq.cond_beta_mmsc}) and (\ref{eq.cond_beta}) are not fulfilled and compare them with the predictions of our Hamiltonian~(\ref{eq.hamiltonian}), which generally applies. 
As an example, we focus on the dependence of the resonances of the coupled \{atom+cavity\} system that occur when the rountrip phase is an integer multiple of $2\pi$. For our model this is the case for
\begin{equation}
Re\{\phi(\Delta_a,\Delta_c)\}=2\pi \times q\;, \label{eq.resonancecondition}
\end{equation}	
with the integer number $q$.

\begin{figure*}
	\includegraphics[width=2\columnwidth]{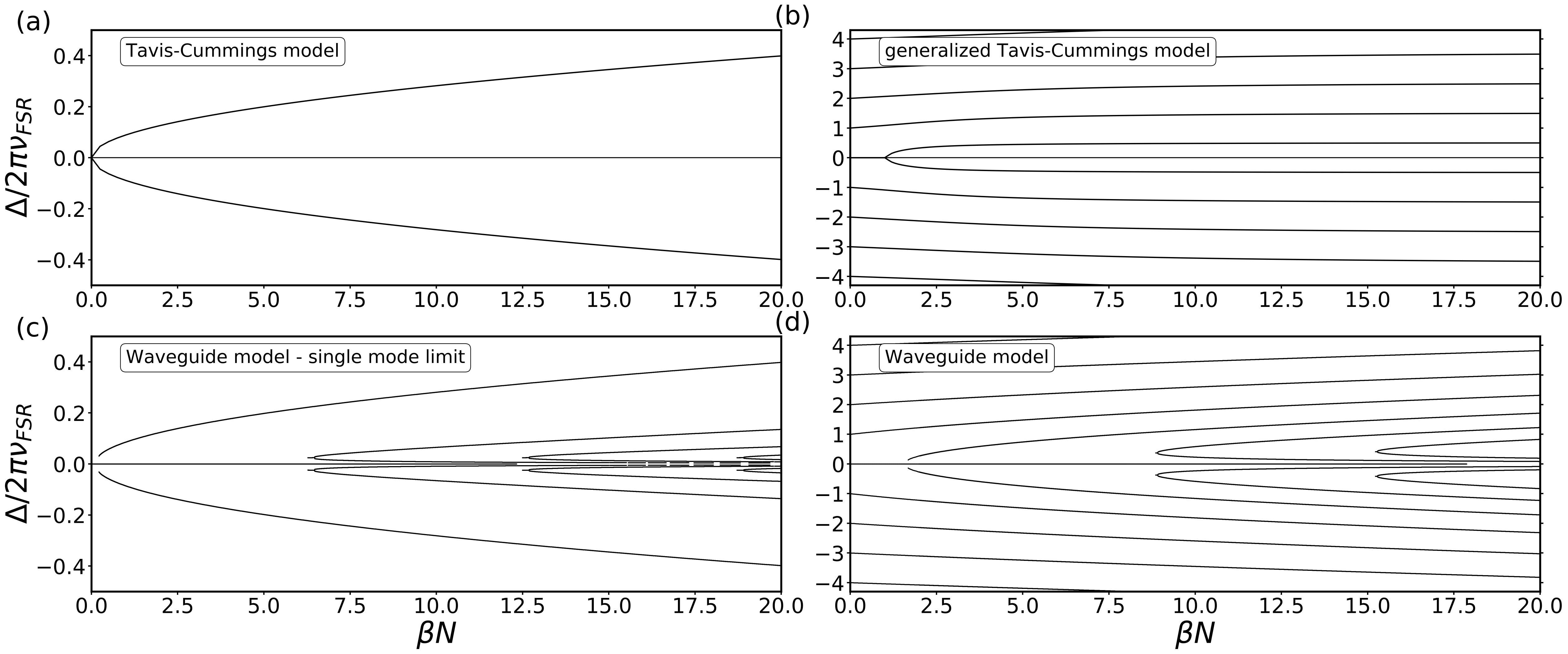}
	\caption{
		Positions $ \Delta = \Delta_{a} = \Delta_c $ of the resonances calculated for the Tavis-Cummings model for (a) $ g_N \ll \nu_{\text{FSR}} $ ($ \nu_{\text{FSR}} = 200$\,MHz, $ \gamma = 2\pi \times 5$\,MHz) and (b) $ g_N \gtrsim \nu_{\text{FSR}} $ ($ \nu_{\text{FSR}} = 10$\,MHz, $ \gamma = 2\pi \times 5$\,MHz) as a function of $ \beta N $. (c) and (d) show the predictions of our model for the same cases and values as in (a) and (b), respectively.  
		In panel (a) we see the usual strong coupling prediction, where the central resonance is split by $ 2g_N \propto \sqrt{N} $. In (b), the system enters the superstrong coupling regime, where the splitting saturates at $\nu_{\text{FSR}}$ and the superstrong coupling-spectrum is similar to the empty cavity spectrum but shifted by $\nu_{\text{FSR}}/2$. 
		In contrast to the predictions of the Tavis-Cummings model, our model ((c) and (d)) predicts the appearance of new resonances for large $\beta N$ in the region around $\Delta = 0$. This is due to the fact that light undergoes multiple $2\pi$ phase shifts in a single resonator roundtrip. Additionally, in (d), where both conditions (\ref{eq.cond1a}) and (\ref{eq.cond1b}) are violated, contrary to the predictions of Tavis-Cummings, the splitting and shifting of the resonances do not saturate but increase continuously with increasing $ \beta N $.
	}	
	\label{fig.posres}
\end{figure*}
\subsubsection{Single-mode regime}
We first consider the case $g_N\ll\nu_{\text{FSR}}$, i.e., the situation where the interaction of the emitter with the higher order cavity resonances can be neglected. Fig.~\ref{fig.posres}(a) and (c) show the predicted cavity resonance as a function of $\beta N$ according to the predictions of the TC model and our model, respectively. For small $\beta N$ in panels (a) and (c), we observe the familiar behavior where the central resonance of the cavity splits into two new resonances that, for $g_N\gg(\kappa,\gamma_l)$, are separated by $2g_N$ and for which the splitting increases with $g_N\propto\sqrt{\beta N}$. At the same time, the adjacent longitudinal resonator modes (not shown) are mostly unaffected.
However, surprisingly, for large atom numbers for which condition~(\ref{eq.cond1b}) or (\ref{eq.cond_beta}) is not fulfilled, the predictions of the two models qualitatively differ. While both models still predict that the splitting of the central line increases $\propto\sqrt{\beta N}$, our model predicts the occurrence of additional resonances in the spectrum close to the atomic resonance.  As $\beta N$ increases, more of these new resonances appear.
 
\begin{figure}[!ht]
		\includegraphics[width=1\columnwidth]{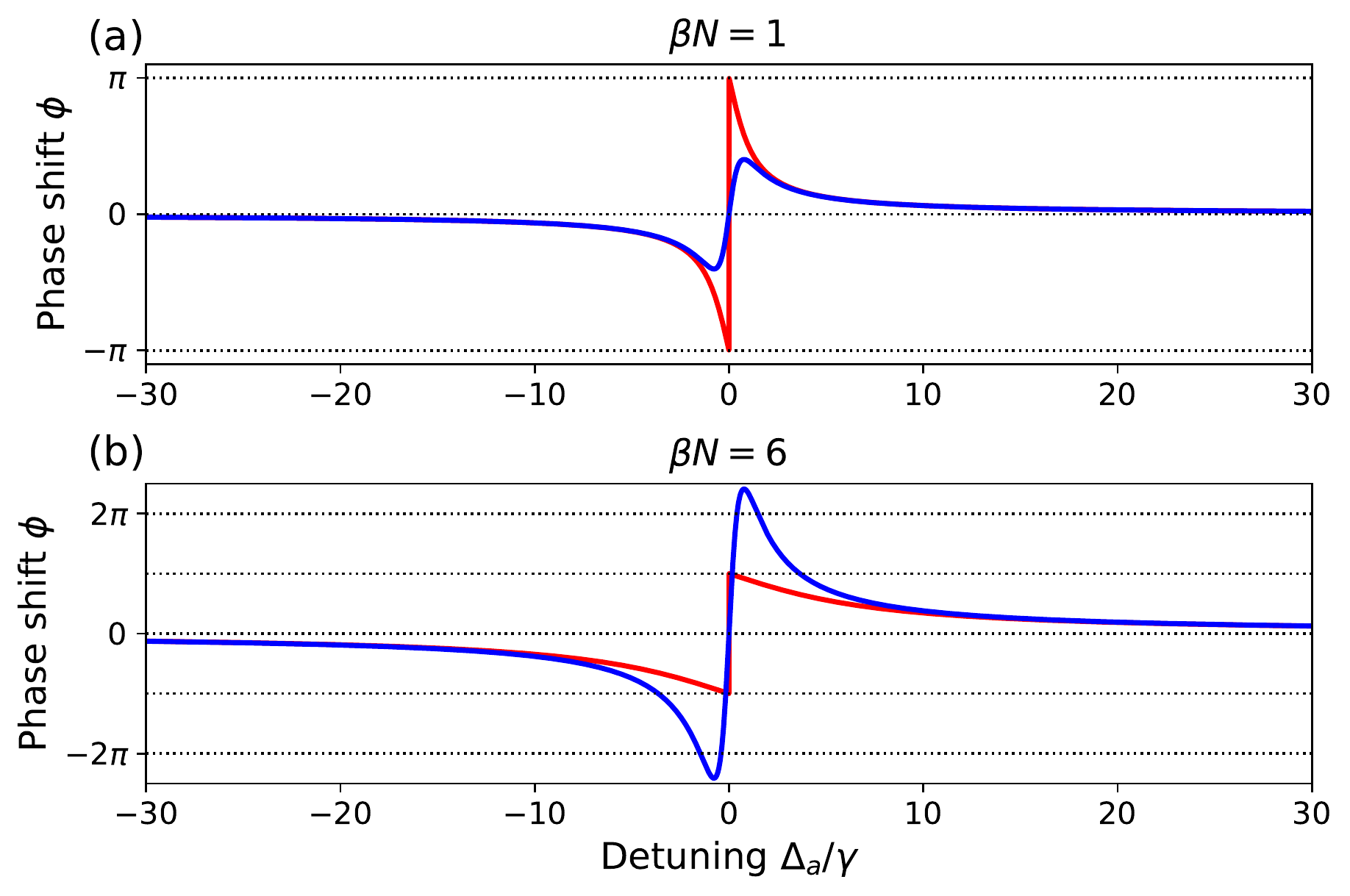}
		\caption{
			Single-pass phase shift of the light induced by a two level quantum emitter as a function of emitter-light detuning $\Delta_a$. (a) The solid red line shows the results for a perfectly coupled single emitter, i.e., $\beta = 1$ and $N=1$, as calculated from Eq.~(\ref{eq.tNDicke}). The blue line represent the predictions of our model from Eq.~(\ref{eq.tAtom}) with $\beta = 0.2$ and $ N = 5 $.
			(b) Same situation as in (a) but now for $\beta N = 6$. The solid red line is the prediction for the Dicke superatom for which the TC model applies. It is calculated from Eq.~(\ref{eq.tNDicke}) for $ \beta = 1, N = 6 $. Interestingly, despite the larger number of emitters, the maximum phase shift of the light imparted by the emitters is not increased compared to (a) and remains within $[-\pi, \pi]$. The blue line ($ \beta = 0.2, N = 30 $) shows the predictions of our model with a maximum phase shift well in excess of $[-\pi, \pi]$.
			}
		\label{fig.shifts}
\end{figure}

A physical explanation of the origin of these additional resonances can be obtained when looking at the roundtrip phase shift of the light  $\mathrm{Re}\left\{ \phi(\Delta_a,\Delta_c) \right\}$. According to Eq.~(\ref{eq.definitionPhi}) this phase shift is comprised of a propagation and an emitter-induced phase shift. For a single emitter as well as the superatom assumed in the TC model, the phase shift imparted by the emitters to the light is always limited to the range $[-\pi,\pi]$, independently of $N$.
However, when coupling a spatially extended ensemble of atoms to the resonator, $t_N$ is given by Eq.~(\ref{eq.t_N}) and the atom-induced single-pass phase shift $\arg(t_N)$, can in principle span an unlimited range and, consequently, more solutions to Eq.~(\ref{eq.resonancecondition}) are found, see Fig.~\ref{fig.shifts}.

\subsubsection{Coupling to more than one mode}
In the limit where $g_N$ exceeds the free spectral range of the resonator, the atoms strongly interact with different resonator modes and, consequently, the TC model has to be replaced by its generalization to more than one mode, as given in Eq.~(\ref{eq.TC_multimode}).
Figure~\ref{fig.posres}(b) shows the cavity resonances predicted by this model as a function of $\beta N$ whereas Fig.~\ref{fig.posres}(d) shows the prediction of our model for the same parameters.
In the case shown in panel (b) one observes that the central resonance starts to split once $g_N$ gets comparable to the loss rate $\gamma_l$ and the system enters the strong coupling regime. With increasing $\beta N$ also the adjacent longitudinal resonances are increasingly shifted outwards. For very large $g_N$, the splitting of the central resonance saturates at $\nu_{\text{FSR}}$ and the shift of all adjacent resonances saturates at $\nu_{\text{FSR}}/2$ with respect to the resonance of the empty resonator. This saturation for large $g_N$ is a consequence of the interaction of the emitter with many modes and can, e.g., be derived from the eigenenergies of the Hamiltonian, see Appendix~\ref{app.MTC}.

In contrast to this, our model does not predict any saturation effect and for $g_N\gg\gamma$ the splitting of the central resonance is simply given by $2g_{N}=2\bar{g}_1\sqrt{N}$, i.e., it follows the $\sqrt{N}$ behavior even when the collective coupling exceeds the free spectral range. The other resonator modes are all subject to a shift proportional to $\sqrt{N}$, for $g\gg \nu_{\text{FSR}}$, and show no saturation effect either. Furthermore, similar as in the single-mode situation, our model predicts the occurrence of new resonances close to the atomic resonance.

In summary, when comparing the resonance frequencies predicted by the TC model and our cascaded interaction model, qualitative differences occur already in the weak driving limit considered here. These differences are present for the single-mode as well as the multimode situation, where in both cases, our model predicts the occurrence of new resonances that are not present in the TC model. Furthermore, for the multimode situation, the TC model predicts a saturation of the shift of the cavity resonances which does not occur in our description.
We note that the new resonances emerge close to or within the atomic absorption profile where the propagating cavity field is subject to strong absorption. As a consequence, they are expected to only give rise to a low-contrast modulation of the transmission signal. This holds in particular for the branches that shift towards the atomic resonance. 

\section{Experimental verification}
\label{sec.experiment}
\begin{figure}[!ht]
	\centerline{\includegraphics[width=1\columnwidth]{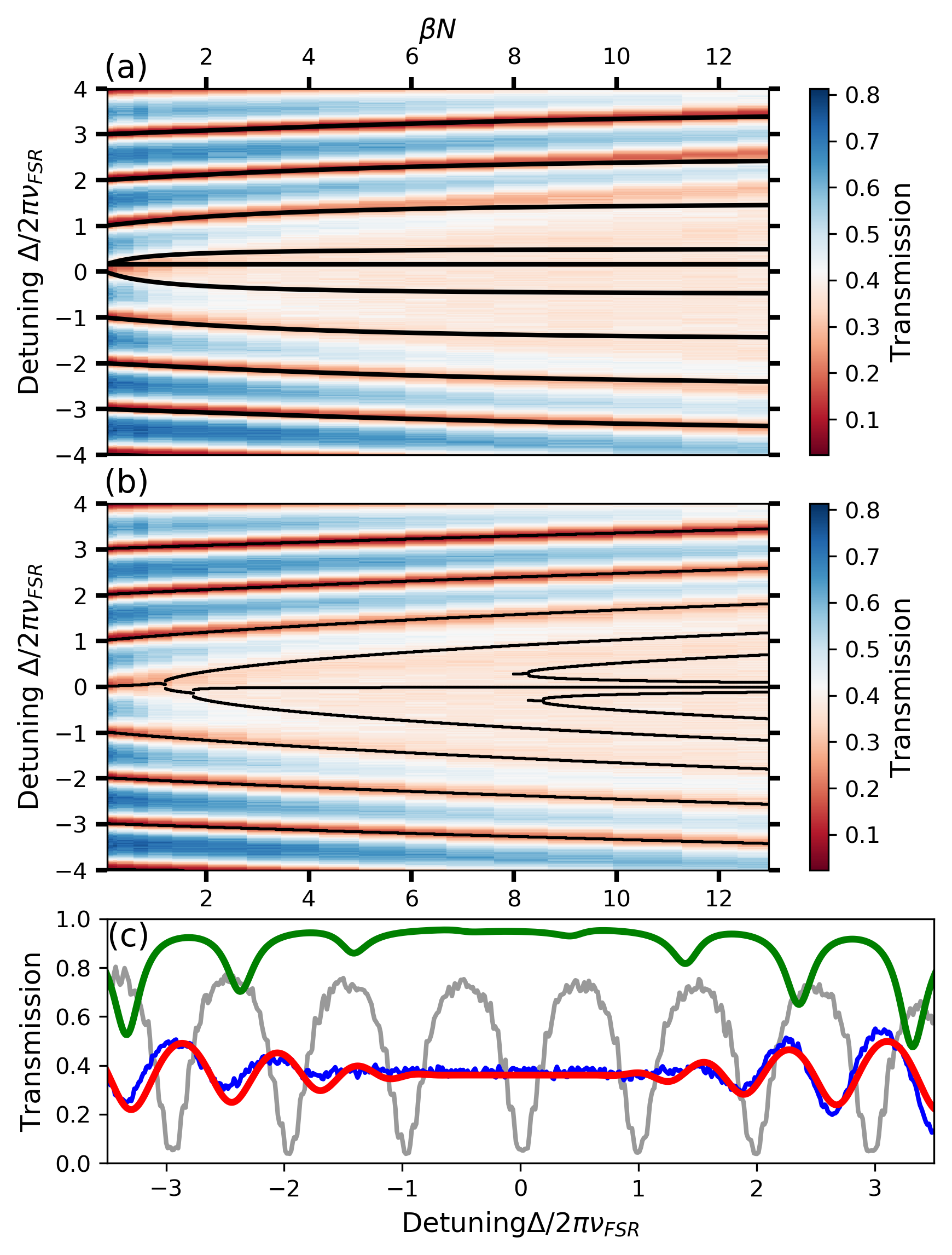}}
	\caption{Experimentally measured transmission spectra of a ring resonator coupled to an optically dense cloud of atoms. The black solid lines indicate the resonance frequencies predicted by the generalized TC model in panel (a) and our model in panel (b), both for $ \gamma/2\pi = 2.61 $\,MHz, $ \beta = 0.005 $ and $ \nu_{\text{FSR}} = 7.1$\,MHz. (c) cut through the data (blue) presented in (a) and (b) for $ \beta N = 12.4 $, with a theory curve using the generalized TC model (green) and a fit to the waveguide model (red), using Eq.~(\ref{eq.solution}). Note that for all theory plots a cavity--atom detuning $ \Delta_{ca} = 1.12$\,MHz is considered, that was present during the measurement. It becomes apparent, that our model agrees well with the experimental data and, for the large coupling strength depicted here, we observe the splitting of the central resonance exceeding the limit of $ \nu_{\text{FSR}} $ and the shift does not saturate, unlike in the TC prediction. Note that, the predicted additional resonances in panel (b) cannot be discerned in the experimental data due to their small contrast and the limited signal to noise of our measurement. The shifting of the first higher resonance exceeds $ \nu_{\textrm{FSR}}/2 $ as can be seen by comparison to the spectrum of the empty resonator (gray), indicating operation in the superstrong coupling regime.
	}
	\label{fig.exp}
\end{figure}

To experimentally illustrate the breakdown of the Tavis-Cummings model, we experimentally investigate the situation where the conditions (\ref{eq.cond1a}) and (\ref{eq.cond1b}) break down using an experimental platform in the superstrong coupling regime, which we reported on in \cite{Johnson2019}. Compared to the measurements presented in this previous publication, we could substantially increase the collective coupling strength, which allows us to now highlight the deviation from the TC model. The experiment consists of a cloud of laser-cooled Cs atoms coupled to a 30-m long ring fiber resonator ($\nu_{\text{FSR}} = 7.1$~MHz) via a nanofiber-based optical interface. The effective number of atoms that are coupled to the resonator mode reaches up to $ \sim 2300 $, thereby giving rise to a maximum $\beta N= 12.96$ (or $g_N=2\pi\times8.74$~MHz). We measure the loaded cavity spectra for a range of different collective coupling strengths, scanning the frequency of the probe field over many free spectral ranges and measuring its transmission with a single photon counter. The result of this measurement is summarized in Fig.~\ref{fig.exp}.

Figures~\ref{fig.exp}(a) and (b) show the measured spectrum together with the predictions of the TC model and our model, respectively. Whereas the position of the cavity resonances do not follow the predictions of the Tavis-Cummings approach, theory and experiment agree well for our model. Specifically, we observe that the shifts of the resonances do not saturate as the collective coupling strength increases and its dependence on $ \beta N $ agrees well with the prediction from our model (solid black lines). Furthermore the maximum shift observed, e.g., for the $ +1\textrm{st}$-order, exceeds $ \nu_{\textrm{FSR}}/2 $ and the fit splitting reveals $ 2g_N \approx 2.3 \nu_{FSR}$. These observations exceed the predictions of the TC approach significantly. We note that, due to the low finesse of the resonator, the predicted contrast of the additional resonances for our experimental settings is only about 0.1\%, which is below our experimental signal to noise ratio. Thus, the new resonances cannot be discerned in our data.

\section{Summary and conclusion}
\label{sc.summary}
In this article, we derived quantitative conditions under which the Tavis-Cummings model fails to correctly describe the interaction of an atomic ensemble with a single optical mode. Our analysis shows that the validity of the Tavis-Cummings model breaks down as soon as the single-pass optical depth of the atomic ensemble approaches $OD\approx 1$. In this case, the collective interaction of the atoms with the resonator mode needs to be accounted for by considering the successive interaction of the propagating cavity field with each atom in the ensemble, because the field is significantly altered by each of these interaction events.
We then presented a more general theoretical model that can be applied for describing cQED in all parameter ranges. Using this model, we analytically derived the steady-state solution in the low power limit and compared it to the predictions of the Tavis-Cummings model. We found qualitative differences between the predictions of the two models concerning, e.g., the transmission spectrum of the coupled atom-cavity system: The Tavis-Cumming model predicts that, irrespectively of the number of coupled atoms, the splitting of the central resonance cannot exceed the resonator's free spectral range. In contrast, our model predicts that the splitting is not limited and continues to grow as the number of atoms increases. Moreover, with increasing atom number, our model predicts a growing number of additional resonances, which are missing by the Tavis-Cummings model. These resonances occur because, for large ensembles, the single-pass phase shift of the cavity field induced by the atoms is in principle not limited. This feature cannot be captured by the Tavis-Cummings model, where the largest possible phase shift is $\pm\pi$. We compare our findings to experimentally measured transmission spectra of a 30-m long optical ring resonator, which is coupled with a high optical depth atomic ensemble. The measured spectra agree well with the predictions of our model and qualitatively disagree with the Tavis-Cummings model.

Our approach is not limited to a concrete realization of the cavity, the quantum emitter, or the emitter-light coupling mechanism. Consequently, our findings are relevant for all theoretical and experimental studies, in which the number of resonator-coupled emitters reaches or exceeds 
$1/(2\beta)$, see Eq.~(\ref{eq.cond_beta2}). In this case, the Tavis-Cummings model should be taken with precaution. This particularly applies to experiments that study superradiance \cite{Araujo2016a,ferioli_storage_2020,solano_super-radiance_2017,Kwong2014} or that make use of cooperative effects to enhance the performance of quantum memories and, thus, invariantly aim for high single-pass optical depths. Adding a cavity with the aim of enhancing the performance then inevitably sets these experiments into a regime where the Tavis-Cummings model cannot be applied. Instead, in general, our approach is required for correctly modeling the system's properties. Furthermore, in the superstrong coupling regime, non-Markovian dynamics has been predicted where, e.g., pulsed revivals rather than conventional Rabi oscillations are expected \cite{krimer_route_2014}.
However, this non-Markovian behavior requires high single-pass optical depths. Thus, the Tavis-Cummings model cannot be expected to hold in this regime while our description is well suited in view of its wide range of applicability. 
We note that Eq.~(\ref{eq.cond_beta2}) can already be violated for a single atom for $\beta\gtrsim1/2$. This shows, that even the single-emitter cavity system can only be described correctly by the JC model for the case of small $\beta$, i.e., for large scattering losses into the environment. Consequently, the case of an atom, perfectly coupled to a cavity mode ($\beta=1$) is not covered by the JC model which will be discussed in \cite{in_preparation}, where we study in detail the situation of a single emitter coupled with high $ \beta $ to a Fabry-P\'erot or ring resonator.

In this manuscript, we limited our discussion to the linear regime of low excitation where at most one photon is present in the cavity. Under these assumptions, we can derive analytical solutions for the system. Already here, we observe qualitative deviations between the Tavis-Cummings approximation and our model. However, one of the main reasons for using optical cavities is the enhancement of non-linear effects. In this context, we also expect to observe qualitative differences concerning the non-linear properties of the system. For example, it has recently been shown theoretically and experimentally that the transmission of light through an atomic ensemble with a large optical depth leads to the generation of photon correlations and squeezing through the collective non-linear response of the atoms \cite{Guimond2016, Mahmoodian2019, prasad_correlating_2020, hinney_unraveling_2020}. The physics underlying these effects is already contained in our Hamiltonian.
This illustrates that the theoretical framework presented in this article has consequences and applications that go well beyond the presented examples. It therefore has the potential to reveal new physics that cannot be captured by the Jaynes- and Tavis-Cummings models.

\section{Acknowledgment}
We acknowledge financial support by the Alexander von Humboldt Foundation in the framework of an Alexander von Humboldt Professorship endowed by the Federal Ministry of Education and Research and by the Austrian Science Fund (NanoFiRe grant project No. P 31115).

\bibliography{HighOD}

\appendix

\section{Strong coupling in the JC (TC) model}
\label{app.strong_coupling}
In order to derive the steady-state of a resonator coupled to a single atom we use a photon transport approach \cite{Shen2009} to quantum mechanically describe the resonator probing fields. The Hamiltonian of this system is given by
\begin{equation}
\label{H1_strong}
\hat{H}= \hat{H}_{JC'}+\hat{U}_{\text{probe}},
\end{equation}
where the first term describes the interaction of a single atom with the cavity field. It includes loss from photons scattered into non-cavity modes by the atom ($ -i\gamma_l $) and loss from the cavity implementation ($ -i\kappa_{0} $) and is given by
\begin{equation}
\hat{H}_{JC'}/\hbar = (\Delta_{a}-i\gamma_l) \hat{\sigma}^+ \hat{\sigma}^- + (\Delta_c-i\kappa_0) \hat{a}^{\dagger} \hat{a} + g(\hat{a}^{\dagger} \hat{\sigma}^- + \hat{a} \hat{\sigma}^+).
\end{equation}
Here, $ \hat{a}^\dagger $ ($ \hat{a} $) is the cavity photon creation (annihilation) operator, $ {\Delta_c=\omega_c-\omega}$ is the cavity--probe detuning, ${\Delta_{a}=\left(\omega_{\mathrm{a}}-\omega\right)}$ is the atom--probe detuning 
and $ g $ the coupling strength.
The probing term describes the coupling of a driving field $ \hat{c}^\dagger_x $ to the cavity via 
\begin{equation}
\frac{\hat{U}_{probe}}{\hbar}=\int_{-\infty}^{\infty}dx\big[ \hat{c}^\dagger_x(\omega_0-iv_g{\partial_x})\hat{c}_x + V_{\text{cav}} \delta(x) ( \hat{a}^\dagger \hat{c}_x + \hat{a} \hat{c}_x^\dagger\ )\big],
\end{equation}
where $ V_{\text{cav}}  = \sqrt{2\kappa_{\text{ext}} v_g}  $ is the coupling strength between cavity and the driving field.
In the weak driving regime we obtain the steady-state solution of $ \hat{H}|\Psi\rangle=E|\Psi\rangle $ by using a general single excitation wavefunction 
\begin{eqnarray}\label{psi1_strong}
|\Psi\rangle&=& \left[\int dx \phi_{\text{c}}(x) \hat{c}_x^\dagger +\phi_{\textrm{cav}} a^\dagger + \phi_{\text{at}} \hat{\sigma}^+\right] |0\rangle.
\end{eqnarray}
The results for the steady-state values for the amplitude of the output (input) field, $ \phi_{o} $ ($\phi_i$), the atomic excitation amplitude, $ \phi_{\text{at}} $, and the cavity field amplitude, $ \phi_{\text{cav}} $, are given by
\begin{eqnarray}
\label{single_atom_strong_transmission}
\frac{\phi_o}{\phi_i}&=& \frac{g^2+(\gamma_l+i\Delta_a)\left(\kappa_{0}-\kappa_{\text{ext}}+i\Delta_c\right)}{g^2+(\gamma_l+i \Delta_a)\left(\kappa_{0}+\kappa_{\text{ext}}+i\Delta_c\right)}\\
\frac{\phi_{\text{at}}}{\phi_i}&=&\frac{-g\sqrt{2v_g\kappa_{\text{ext}}}}{g^2+(\gamma_l+i\Delta_a)(\kappa_{0}+\kappa_{\text{ext}}+i\Delta_c)}\\
\frac{\phi_{\text{cav}}}{\phi_i}&=&\frac{-i\sqrt{2v_g\kappa_{\text{ext}}}(\gamma_l+i\Delta_a)}{g^2+(\gamma_l+i\Delta_a)(\kappa_{0}+\kappa_{\text{ext}}+i\Delta_c)}\\
\frac{\phi_{\text{at}}}{\phi_\text{cav}}&=&\frac{-ig}{\gamma_l + i\Delta_a},
\end{eqnarray}
where $ |\frac{\phi_{\text{at}}}{\phi_\text{cav}}|^2= \frac{g^2}{\gamma_l^2 + \Delta_a^2}$ is the cavity-induced atomic excitation probability, such that the atom-induced loss from the cavity on resonance can be calculated to be $g^2/\gamma_l$, appearing in cond.~(\ref{eq.cond1b}). The results above apply for both, the JC and TC Hamiltonian with $ g = g_1 $ or $ g = g_N $.

\section{Chiral coupling to a ring cavity}
\label{app.solution}
Here we outline the derivation of Eq.~(\ref{eq.solution}) in the main text. For chiral interaction, the Hamiltonian in Eq.~(\ref{eq.hamiltonian}) can be simplified by neglecting the resonator mode $ \hat{a}^\dagger_{l_-} $ ($ \hat{a}_{l_-} $) that is not coupled to the atoms ($ V_{n-} = 0 $) and consequently $ \hat{c}^\dagger_-, \hat{c}_- = 0 $ and $ \hat{d}^\dagger_-, \hat{d}_- = 0$. This gives
\begin{equation}
\label{eq.hamiltonian_app}
\begin{aligned}
&\hat H/\hbar=\hat{U}_{c}+ \int_{-\infty}^{0}dx \left[ \hat{c}_{x_+}^{\dagger}\left(\omega-i v_{g} \partial_x \right) \hat{c}_{x_+} \right]\\
& + \qquad\int_{0}^{+\infty} dx \left[\hat{d}_{x_+}^{\dagger}\left(\omega-i v_{g} \partial_x \right) \hat{d}_{x_+} \right] \\
& + \qquad\int_{0}^{L} dl\Big\lbrace\left[\hat{a}^{\dagger}_{l_+}\left(\omega+i v_{c} {\partial_l}\right) \hat{a}_{l_+}\right]\\
&+\qquad\sum_{n=1}^{N}\delta(l-l_n)\Big[\hat{\sigma}_{n}^{+} \hat{\sigma}_{n}^{-}\left(\omega_{\mathrm{a}}-i\gamma_l\right)\\
&+\qquad V_{n+}\left(\hat{\sigma}_{n}^{+} \hat{a}_{l_+}+\hat{\sigma}_{n}^{-} \hat{a}_{l_+}^{\dagger}\right)\Big]\Big\rbrace.
\end{aligned}
\end{equation}
The general single excitation wavefunction of the system is of the form
\begin{equation}\label{app.wavefunction}
\begin{aligned}
|\Psi\rangle=&\left[\int_{-\infty}^{0}dx \phi_c(x)\hat{c}_{x_+}^\dagger + \int_{0}^{\infty}dx\phi_d(x)\hat{d}_{x_+}^\dagger \right. \\
&\left. +\int_{0}^{L}dl \phi_a(l)\hat{a}_{l_+}^\dagger + \sum_{n=1}^{N}\phi_{\textrm{at},n} \hat{\sigma}^+_n\right] |0\rangle.
\end{aligned}
\end{equation}
To obtain the steady state solution we make the ansatz that the fields can be described as propagating plane waves with wavenumber $ k $ of the form
\begin{equation}
\begin{aligned}
\phi_{c}(x) &=e^{i k x} \phi_{c} \Theta(-x)\\
\phi_{d}(x)&=e^{i k x} \phi_{d} \Theta(x)\\
\phi_{a}(l) &=e^{-i k l}\sum\limits_{n=0}^{N}\phi_n\Theta(l-l_n)\Theta(l_{n+1}-l),
\end{aligned}
\end{equation}
where $ l_0=0 $ and $ l_{N+1} = L $, $\Theta$ is the Heaviside step function and $ \phi_{c} $ and $ \phi_{d} $ are complex numbers. Here, the $n$th atom couples to the resonator at position $ l_n $ with $ \phi_n $ being the cavity field after the $ n $th atom and $ \phi_{\textrm{at},n} $ the excitation amplitude of atom $ n $.
For simplicity, we set $ v_g = v_c $.
Using these relations, one can unambiguously solve the eigenvalue problem  $H|\Psi\rangle=E|\Psi\rangle$.
Injecting the wavefunction ansatz into the Schr\"odinger equation we obtain a set of coupled equations
\begin{equation}
\begin{aligned}
0&=-i v_g \frac{\phi_{d}}{2}-v_g t_{\mathrm{rt}}t \frac{\phi_{N}}{2} e^{-i k L}+i v_g r \frac{\phi_c}{2}\\
0&=i v_g \frac{\phi_{0}}{2}-i v_g t_{\mathrm{rt}}r \frac{\phi_{N}}{2} e^{-i k L}+v_g t \frac{\phi_c}{2}\\
0&=-iv_g\left(\phi_{n}-\phi_{n-1}\right)e^{-ikl_n}+V_{n+}\phi_{\textrm{at},n} \\
0&=\frac{V_{n+}}{2}\,\left(\phi_{n}+\phi_{n-1}\right)e^{-ikl_n}+\left(\Delta_a-i\gamma_l\right)\phi_{\textrm{at},n}.
\end{aligned}
\end{equation}
For the waveguide atom coupling 
$ V_{n+} = \sqrt{2\beta_{n}\gamma v_g} $ and the atom--field detuning $ \Delta_{a} = \left(\omega_{\mathrm{a}}-\omega\right) $ we can solve the set of equations above. For the resonator fields we get
\begin{equation}\label{app.tat}
\begin{aligned}
\frac{\phi_{n}}{\phi_{n-1}}&= 1-\frac{2\beta_{+,n}\gamma}{\gamma+i\Delta_{a}} = t_{n}\;\;\; n\neq 0\\
\frac{\phi_0}{\phi_c}&=\frac{-i t }{e^{-ikL}rt_Nt_\textrm{rt}-1}.
\end{aligned}
\end{equation}

For the field at the output of the probed atom-cavity system we get 
\begin{equation}
\frac{\phi_d}{\phi_c}=\frac{t_{N} t_\textrm{rt}	e^{-ikL}-r}{t_{N} t_\textrm{rt}re^{-ikL}-1}.
\end{equation}
Using $ kL = \Delta_c/\nu_{\text{FSR}} $ we obtain for the power reflection of a loaded ring resonator Eq.~(\ref{eq.solution})
\begin{equation}
R =\left|\frac{\phi_d}{\phi_c}\right|^2 =  \left|\frac{e^{-i\Delta_c/\nu_{\text{FSR}}}t_{\text{rt}}t_N-r}{e^{-i\Delta_c/\nu_{\text{FSR}}}t_{\text{rt}}t_N r-1}\right|^2.
\end{equation}
Apart from the resonator fields between the atoms in Eq.~(\ref{app.tat}), all quantities only depend on the total single-pass transmission $t_N$ through the whole atomic ensemble, which is given by
\begin{eqnarray}
t_N&=&\prod_{n=1}^N t_{n}
\end{eqnarray} 
For $\beta_{n}$  sufficiently small or sufficiently small variations of $\beta_n$ we can approximate $t_N$ via
\begin{eqnarray}
t_N&=&1-\sum_n \tilde\beta_n+\sum_{n<m}\tilde\beta_n\tilde\beta_m- ...\nonumber \\
&=& 1-\underbrace{\sum_n \tilde\beta_n}_{=N\tilde\beta}+\underbrace{\big(\sum_n \tilde\beta_n\big)^2 -\sum_n\tilde\beta_n^2}_{\approx\frac{1}{2}(N^2-N)\beta^2}- ... \nonumber \\
&\approx&\sum_n\left(\begin{array}{c}
N\\n
\end{array}\right)(-\tilde\beta)^n=(1-\tilde\beta)^N,
\end{eqnarray} 
where we used the shorthand notation $\tilde\beta_n=2\beta_{n}\gamma/(\gamma+i\Delta_a)$ and $\tilde\beta=2\beta\gamma/(\gamma+i\Delta_a)$, with $\beta=\sum\beta_{n}/N$ describing the mean atom-mode coupling. We note that for large detuning $\Delta_a>\gamma$, this approximation is always fulfilled. 

\section{Coupling strength g and $\beta$-factor}
\label{app.g_to_beta}
To compare the description of our model to the Jaynes- and Tavis-Cummings models, we need an expression linking the characteristic coupling parameters $ g_1 $ and $ \beta $.
With the definition $ \langle \hat{a}^\dagger \hat{a} \rangle = \langle \hat{n} \rangle $, where $ \hat{n} $ is the number operator quantifying the mean number of photons inside the cavity and $ \hat{a}^\dagger $ ($ \hat{a} $) the photon creation (annihilation) operator in the Jaynes-Cummings Hamiltonian. For our model the mean intracavity photon number is $ \langle \int_{0}^{L} dl \, \hat{a}_l^\dagger \hat{a}_l \rangle = \langle \hat{n} \rangle $. With this we can identify $ V_{n+}^2/L = g^2 $, which leads to
\begin{equation}
g^2 = 4\beta\gamma\nu_{\textrm{FSR}}
\end{equation}
for an emitter placed inside an anti-node of the cavity field in a Fabry-P\'erot cavity (symmetric coupling) and 
\begin{equation}
g^2 = 2\beta\gamma\nu_{\textrm{FSR}}
\end{equation}
for a ring cavity, where we assumed chiral coupling.

\section{Solutions of our cascaded model for large $ \nu_{\text{FSR}}$}
\label{app.comptransm}
From the transmission in the cascaded real-space formalism in Eq.~(\ref{eq.solution}), one can reproduce the transmission spectra calculated from the Jaynes- and Tavis-Cummings model for a resonator coupled strongly to atoms, for $ \beta N < 1 $. At first we substitute for $ \beta_n\gamma $ using $ g^2=2\beta_n\gamma\nu_{\textrm{FSR}} $ as 
\begin{equation}
t_{n} = 1-\frac{2 \beta_n \gamma}{\gamma+i \Delta_a} =
1-\frac{2 g^2}{g^2 +2 \nu_{\textrm{FSR}}(\gamma_l+i \Delta_a)}.
\end{equation}
For single-mode interaction, where $ \Delta_c \ll \nu_{\text{FSR}} $, a condition satisfied in most conventional cQED experiments, we can restrict the derivation to first order in $ \nu_{\textrm{FSR}} $ and obtain for the reflection from the incoupling mirror
\begin{equation}
R= \left|\frac{g^{2}+(\gamma_l+i\Delta_a)\left(\kappa_{0}-\kappa_{\text{ext}}+i\Delta_c\right)}{g^2+(\gamma_l+i \Delta_a)\left(\kappa_{0}+\kappa_{\text{ext}}+i\Delta_c\right)}\right|^2,
\end{equation}
the same expression as from Jaynes- and Tavis-Cummings as derived in Eq.~(\ref{single_atom_strong_transmission}) for $ g = g_1 $ or $ g = g_N $, respectively.

\section{Multimode extensions of the Jaynes- and Tavis-Cummings models}
\label{app.MTC}
Starting from the standard Jaynes-Cummings approach the generalization to many modes is 
\begin{equation}
\frac{\hat{H}}{\hbar} = \omega_{\textrm{at}}\hat{\sigma}_+\hat{\sigma}_- + \sum_j \left[\omega_j\hat{a}^\dagger_j\hat{a}_j+ g_j\left(\hat{a}^\dagger_j\hat{\sigma}_-+\hat{a}_j\hat{\sigma}_+\right)\right],
\end{equation}
where $ \hat{a}^\dagger_j $ creates a photon in the $ j $th mode of the resonator and $g_j$ is the coupling strength between the emitter and the cavity mode $j$.
In a similar way, we can apply the same method to the Tavis-Cummings model
\begin{equation}
\begin{aligned}
\frac{\hat{H}}{\hbar} =& \omega_{\textrm{at}}\hat{S}^+\hat{S}^-\\
&+ \sum_j \left[\omega_j\hat{a}^\dagger_j\hat{a}_j +  g_{N,j}\left(\hat{a}^\dagger_j\hat{S}^-+\hat{a}_j\hat{S}^+\right)\right],
\end{aligned}
\end{equation}
where we substituted the operators $ \hat{S}^\pm = 1/\sqrt{N}\sum_n\hat{\sigma}^\pm_n $, to take a collective atom-resonator interaction into account. 
In order to calculate the eigenvalue spectrum of the Hamiltonian, we now  consider an equally distributed mode spacing of the resonator around the atomic resonance as $ \omega_j = j\omega_{\text{FSR}} + \Delta_a $, where we define $ \omega_0 $ as the resonance closest to the atomic resonance. The resonator modes are separated by the free spectral range $ \omega_{\textrm{FSR}} = 2\pi \nu_{\textrm{FSR}} $ and $ \Delta_{a} = \omega_{\mathrm{at}}-\omega_0$ is the atom--resonator detuning. Furthermore, for simplicity, we assume that the atoms couple with equal strength to all cavity modes with $g=g_{N,j}$ which is, e.g., the case for a ring resonator.
In order to solve the Schr\"odinger equation $ H|\Psi\rangle = \hbar \omega |\Psi\rangle $, we make a single-excitation wavefunction ansatz $ |\Psi\rangle = \sum_j \alpha_j \hat{a}^\dagger_j + \beta \hat{\sigma}^+|0\rangle$.
Plugging the ansatz into the time-independent Schr\"odinger equation and comparing coefficients leads to
\begin{eqnarray}
\alpha_j &=&\frac{g\beta}{\omega-j\omega_{\text{FSR}}-\Delta_{a}} \;\; \textrm{and}\\
\beta &=& \frac{g}{\omega}\sum_{j=1}^N\alpha_j.
\end{eqnarray}
From this, we can formulate an eigenvalue equation
\begin{equation}
1 = \frac{g^2}{\omega}\sum_{j=-M}^M\frac{1}{\omega-j\,\omega_{\textrm{FSR}}-\Delta_{a}},
\end{equation}
having $ 2(M+1) $ solutions, where $ 2M+1 $ is the number of modes considered.
In the limiting case of very large coupling strength $ g\gg\omega_{\text{FSR}} $, we can make an ansatz for $ \omega = m\omega_{\text{FSR}} $ ($ m\in \mathbb{R} $) and approximate $ \omega_{\textrm{FSR}}^2/g^2 \rightarrow 0 $. With this we arrive at
\begin{equation}
0 = \sum_{j=-M}^{M}\frac{1}{m-(j+\Delta_{a}/\nu_{\textrm{FSR}})}.
\end{equation}
For $M\rightarrow\infty$ this equation can be solved for any mode $k$ ($ k \in \mathbb{N} $) by $ m_k = \frac{1}{2}+\Delta_{a}/\omega_{\textrm{FSR}}+k$. Consequently, for a coupled multimode atom--resonator system we find the position of the new resonances of the \{atom + cavity\} system at the frequencies
\begin{equation}
\tilde{\omega}_k = \omega_k+\frac{1}{2}\omega_{\textrm{FSR}},
\end{equation}
where $ k $ is the resonance's mode-number and $\omega_k$ the frequency of the bare cavity resonance. All resonances are shifted by $ \omega_{\textrm{FSR}}/2 $ from their position of the bare cavity system, where the shift saturates.
\end{document}